\newcommand{\version}{May 15, 2023}
\documentclass[letterpaper,oneside,11pt,pdftex]{article}
\pdfoutput=1
\usepackage[bindingoffset=0.0cm,textheight=22.60cm,hdivide={*,16.25cm,*}, vdivide={*,22.60cm,*}]{geometry}
\usepackage{amsmath,amsfonts,amssymb}
\usepackage{mathtools}
\usepackage[pdftex]{graphicx}
\usepackage[margin=1cm,font=small]{caption}
\usepackage[T1]{fontenc}
\usepackage[utf8]{inputenc}
\usepackage{fouriernc}

\usepackage[numbers,square,comma,sort&compress]{natbib}
\usepackage{import}

\usepackage[pdftex,hyperref,svgnames]{xcolor}
\usepackage[pdftex,bookmarksnumbered=true,breaklinks=true,%
colorlinks=true,linktocpage=true,linkcolor=MediumBlue,citecolor=ForestGreen,urlcolor=DarkRed]{hyperref}

\makeatletter
\AtBeginDocument{
  \hypersetup{
    pdfkeywords = {\keywords},
    pdftitle = {\@title},
    pdfauthor = {\@author}
  }
}
\makeatother

\makeatletter

 \@addtoreset{equation}{section}
 \makeatother

  \newcommand{\cL}{\mathcal{L}}

 \newcommand{\cT}{\mathcal{T}}

\newcommand{\pa}{\partial}

\newcommand{\sgn}[1]{\textrm{sgn}\!\left(#1\right)}
\newcommand{\nn}{\nonumber}
\newcommand{\eqnref}[1]{Eq. \eqref{#1}}

\newcommand{\ct}{c_{\textrm{T}}}
\newcommand{\cl}{c_{\textrm{L}}}
\newcommand{\gt}{\gamma_{\textrm{T}}}
\newcommand{\gl}{\gamma_{\textrm{L}}}
\newcommand{\bt}{\beta_{\textrm{T}}}

\renewcommand{\Im}{\mathrm{Im}}

\newcommand{\txt}[1]{\textrm{#1}}

\DeclarePairedDelimiter\abs{\lvert}{\rvert}

\title{\texorpdfstring{\begin{flushright}
        {\small LA-UR-22-22567}
       \end{flushright}\vspace{2em}}{}%
       Properties of accelerating edge dislocations in arbitrary slip systems with reflection symmetry}

\author{Daniel N. Blaschke, Khanh Dang, Saryu Fensin, Darby J. Luscher}

\date{\version}

\newcommand{\keywords}{dislocations in crystals, dislocation mobility, crystal plasticity, transsonic motion}

\graphicspath{{./}{./figs/}}

\begin{document}

 \maketitle

 \thispagestyle{empty}
 \begin{center}
 \vspace{-0.3cm}
 Los Alamos National Laboratory, Los Alamos, NM, 87545, USA
 \\[0.5cm]
 \ttfamily{E-mail: dblaschke@lanl.gov, kqdang@lanl.gov, saryuj@lanl.gov, djl@lanl.gov}
 \end{center}

\begin{abstract}
We discuss the theoretical solution to the differential equations governing accelerating edge dislocations in anisotropic crystals.
This is an important prerequisite to understanding high speed dislocation motion, including an open question about the existence of transonic dislocation speeds, and subsequently high rate plastic deformation in metals and other crystals.
\end{abstract}

\tableofcontents

\section{Introduction and background}
\label{sec:intro}

Dislocations can influence many materials' properties such as thermal conductivity \cite{Termentzidis:2018}, thermal stability \cite{Li:2023}, impact sensitivity \cite{Darku:2022}, ferroelectricity \cite{Lubk:2013}, and electrical resistance \cite{Szot:2006}.
At extremely high rates, plastic deformation is governed by high speed dislocations, a regime where dislocation mobility is poorly understood  \cite{Hansen:2013,Luscher:2016,Blaschke:2019a}.
High speed dislocations experience a drag force due to scattering phonons (an effect known as `phonon wind') and this interaction (and thus dislocation mobility) is sensitive to the stress distribution in the vicinity of the moving dislocation.
Dislocation drag is thus key to predicting  material strength at extremely high stress and understanding high rate plastic deformation \cite{Gurrutxaga:2020}.
The first principles phonon wind theory was derived in the isotropic and steady state limit for dislocation glide velocities that are much smaller than the transverse sound speed some time ago, see the excellent review article \cite{Alshits:1992}.
More recently dislocation drag theory was generalized to very high (but still subsonic) dislocation velocities \cite{Blaschke:2019Bpap} and anisotropic crystals
\cite{Blaschke:2018anis}, though the effects of acceleration have so far been neglected.

Another key question in this regard is whether dislocations can reach transonic and supersonic speeds under sufficiently high stress.
The only indication that such speeds are possible comes from molecular dynamics (MD) simulations \cite{Olmsted:2005,Marian:2006,Tsuzuki:2008,Oren:2017,Peng:2019,Blaschke:2020MD,Dang:2022Mg}.
Experiments cannot track dislocations in real time at these high speeds\footnote{
After the present manuscript was completed, Ref. \cite{Katagiri:2023} appeared which for the first time measured transonic dislocations in diamond in real time.%
}, but one can hope to indirectly determine the presence of supersonic dislocations and perhaps estimate the fraction and velocity of these dislocations in the near future \cite{Wehrenberg:2017,Dresselhaus:2021}.
This in turn requires a thorough understanding of the solutions to the differential equations governing dislocations, i.e. the equations of motion supplemented by the (leading order) stress-strain relations.

Dislocation theory predicts divergences in self energy and stress at certain limiting velocities \cite{Blaschke:2021vcrit,Teutonico:1961, Teutonico:1962, Barnett:1973b} for steady state dislocations.
In the isotropic limit, it was shown \cite{Markenscoff:2008,Markenscoff:2009,Huang:2009} that an acceleration term together with a regularized dislocation core removes the divergence, thereby opening the possibility of supersonic events.
Other authors emphasized the importance of size variations of the dislocation core as a function of dislocation velocity \cite{Pillon:2007,Pellegrini:2010,Pellegrini:2014,Pellegrini:2020}.
The steady-state solution for dislocations in arbitrary anisotropic crystals has been known for some time \cite{Bacon:1980,Pellegrini:2017}.
The case of accelerating dislocations in anisotropic crystals has also been studied \cite{Markenscoff:1984JE,Markenscoff:1984,Markenscoff:1985a,Markenscoff:1987,Payton:1985,Payton:1995,Blaschke:2020acc}, with pure screw dislocations having been discussed in the most detail \cite{Markenscoff:1984JE,Blaschke:2020acc}.
The most general solution has been given only in a very formal form \cite{Markenscoff:1987}, apart from an additional asymptotic wave front analysis.
In this paper, we consider formal derivation of the accelerating dislocation field of Ref. \cite{Markenscoff:1987} as a starting point to discuss in detail the solution of an accelerating pure edge dislocation in anisotropic crystals.

In particular, we discuss the solution to the following set of differential equations for accelerating dislocations for the special case of pure edge dislocations:
\begin{align}
	\partial_i \sigma_{ij}  &= \rho \ddot{u}_j
	\,, &
	\sigma_{ij} &= C'_{ijkl} u_{k,l}
	\label{eq:diffeqns1}
\end{align}
in coordinates aligned with the dislocations, i.e. $\hat z$ is aligned with the dislocation line and $\hat y$ is parallel to the slip plane normal.
The components of the tensor of second order elastic constants (SOEC) is always measured in Cartesian coordinates that are aligned with the crystal axes, and thus this tensor must be rotated into our present coordinate basis, i.e.:
\begin{align}
	C'_{ijkl} = U_{ii'}U_{jj'}U_{kk'}U_{ll'}C_{i'j'k'l'}
\end{align}
with rotation matrix $U$.

In order to study pure edge (or pure screw) dislocations, the rotated tensor of SOEC must fulfill the following symmetry requirements (shown here in Voigt notation which maps index pairs to single digits, $(11, 22, 33, 32/23, 31/13, 21/12) \rightarrow (1, 2, 3, 4, 5, 6)$):
\begin{align}
	C'_{ij} = \left(\begin{array}{cccccc}
		c'_{11} & c'_{12} & c'_{13} & 0 & 0 & c'_{16} \\
		c'_{12} & c'_{22} & c'_{23} & 0 & 0 & c'_{26} \\
		c'_{13} & c'_{23} & c'_{33} & c'_{34} & c'_{35} & c'_{36} \\
		0 & 0 & c'_{34} & c'_{44} & c'_{45} & 0 \\
		0 & 0 & c'_{35} & c'_{45} & c'_{55} & 0 \\
		c'_{16} & c'_{26} & c'_{36} & 0 & 0 & c'_{66}
	\end{array}
	\right)
	,\label{eq:SOECreflectionplane}
\end{align}
i.e. the six components $c'_{14}$, $c'_{15}$, $c'_{24}$, $c'_{25}$, $c'_{46}$, and $c'_{56}$ must vanish, see Refs. \cite{Foreman:1955} and \cite[Sec. 13-4]{Hirth:1982}.
This ensures that $u_3=0$ implies $\partial_i\sigma_{i3}=0$, and likewise that $u_1=0=u_2$ implies $\partial_i \sigma_{i1}=0=\partial_i \sigma_{i2}$,
so that there exists a $u_3$ that solves the differential equations independently from the pair $(u_1,u_2)$ and vice versa.
Note that in the present coordinates, $u_i$ can only depend on $x$, $y$, and $t$, but not on $z$.
This latter property implies that non-vanishing components $c'_{34}$ and $c'_{35}$ are allowed since they do not enter the differential equations above for pure screw or pure edge dislocations.
On the other hand, the stronger condition $c'_{34}=0=c'_{35}$ implies that the $x_1$, $x_2$ plane is a reflection plane (and then $\sigma_{33}=0$ for pure screw dislocations rather than the weaker $\partial_3\sigma_{33}=0$).

The most general solution for pure screw dislocations was recently derived in Ref. \cite{Blaschke:2020acc}.
The case of accelerating pure edge dislocations was previously studied by Markenscoff and Ni for the special case of $c'_{16}=0=c'_{26}$ (in addition to \eqref{eq:SOECreflectionplane}) in Refs. \cite{Markenscoff:1984,Markenscoff:1985a}, and the general case was presented in Ref. \cite{Markenscoff:1987}.
In Refs. \cite{Markenscoff:1984,Markenscoff:1987}, only a formal solution was derived, though not in closed form.
Here, we present for the first time, a numerical implementation of the accelerating dislocation field for pure edge dislocations in various anisotropic slip systems and study its properties.
Our code is included in version 1.2.7 of PyDislocDyn \cite{pydislocdyn}.

\section{Most general differential equations for pure edge dislocations}

Following Ref. \cite{Markenscoff:1987} in this subsection, but setting $u_3=0$ and plugging the most general rotated tensor of SOEC fulfilling the required properties for studying pure edge dislocations, Eq. \eqref{eq:SOECreflectionplane}, into the differential equations \eqref{eq:diffeqns1}, we find:
\begin{subequations}
\begin{align}
\rho \ddot{u}_1 &= \left(c_{11}\pa_1^2 + 2c_{16}\pa_1\pa_2 + c_{66}\pa_2^2\right)u_1
+ \left(c_{16}\pa_1^2 + \left(c_{12}+c_{66}\right)\pa_1\pa_2 + c_{26}\pa_2^2\right)u_2
\,,
\end{align}
\begin{align}
\rho\ddot{u}_2 & = \left(c_{66}\pa_1^2 + 2c_{26}\pa_1\pa_2 + c_{22}\pa_2^2\right)u_2
+ \left(c_{16}\pa_1^2+ \left(c_{12}+c_{66}\right)\pa_1\pa_2 + c_{26}\pa_2^2\right)u_1
\,.
\end{align}
\end{subequations}
Note that we have dropped the primes on the elastic constants for notational simplicity; nonetheless all $c_{ij}$ are understood to be in the rotated frame aligned with the edge dislocation.
Additionally, we have the boundary conditions
\begin{subequations}
\begin{align}
\lim_{y\to0^\pm}u_1(x,y,t) &= \pm\frac{b}{2}\Theta(x-l(t))
\,,\qquad\forall t>0
\,,\label{eq:bc_isoedge}
\end{align}
\begin{align}
\lim\limits_{y\to0}\sigma_{22}
=\lim\limits_{y\to0}\left(c_{12}\pa_1u_1 + c_{22}\pa_2u_2 + c_{26}\left(\pa_1u_2+\pa_2u_1\right)\right)
=0\,,
\end{align}
\end{subequations}
where $\Theta(x)$ denotes the Heaviside step function, $b$ is the Burgers vector length, and the slip plane is located at $y=0$.
Clearly, the above differential equations and their boundary conditions simplify significantly when $c_{16}=0=c_{26}$, which is what was studied in Refs. \cite{Markenscoff:1984,Markenscoff:1985a}.

In order to solve these more general equations, we apply a Laplace transform in time, i.e.
\begin{align}
{\cal L}\{u_i\}(s) &= \int_0^\infty u_i(t) e^{-st} dt
\,, \label{eq:LaplaceTrafo}
\end{align}
as well as a two-sided Laplace transform (which is related to the Fourier transform with $s\lambda\to ik$) in $x$, i.e.
\begin{align}
{\cal T}\{u_i\}(\lambda) &= \int_{-\infty}^{\infty} u_i(x) e^{s\lambda x} dx
\,, \label{eq:FourierTrafo}
\end{align}
and thus
$U_i(\lambda,y,s)\equiv {\cal T}\{{\cal L}\{u_i(x,y,t)\}\}$.
The transformed differential equations read
\begin{subequations}\label{eq:diffeq_transformed}
\begin{align}
\rho s^2 U_1 &= \left(s^2\lambda^2 c_{11} - 2s\lambda c_{16}\pa_2 + c_{66}\pa_2^2\right)U_1
+ \left(s^2\lambda^2 c_{16} - s\lambda\left(c_{12}+c_{66}\right)\pa_2 + c_{26}\pa_2^2\right)U_2
\,,
\end{align}
\begin{align}
\rho s^2 U_2 & = \left(s^2\lambda^2c_{66} - 2s\lambda c_{26}\pa_2 + c_{22}\pa_2^2\right)U_2
+ \left(s^2\lambda^2c_{16} - s\lambda\left(c_{12}+c_{66}\right)\pa_2 + c_{26}\pa_2^2\right)U_1
\,.
\end{align}
\end{subequations}
Likewise, the transformed boundary conditions in the upper half plane ($y\ge0$) read
\begin{subequations}
\begin{align}
\lim_{y\to0^+}U_1(\lambda,y,s) &= \frac{b}{2s}\int_0^\infty e^{s\lambda x'}\left(1-e^{-s\eta(x')}\right)dx'
\,,\label{eq:transformedbound1a}
\end{align}
\begin{align}
\lim\limits_{y\to0^+}\left(-s\lambda c_{12}U_1 + c_{22}\pa_2U_2 + c_{26}\left(-s\lambda U_2+\pa_2U_1\right)\right)
=0\,,
\end{align}
\end{subequations}
where $\eta(x)\equiv l^{-1}(x)$ and the integral over time was carried out explicitly as described in Ref. \cite{Blaschke:2020acc}.
Additionally, we demand $\lim\limits_{y\to\infty}U_i=0=\lim\limits_{y\to\infty}\pa_2U_i$.
Markenscoff \cite{Markenscoff:1987} argues that the problem can be reduced to a problem on a half-space, so that we now assume $y\ge0$ in the following derivation, and we will generalize to negative $y$ only at the very end.
Note that the first term in boundary condition \eqref{eq:transformedbound1a} is identified as that of the static problem which cannot be treated by a Laplace transform without running into convergence issues \cite{Markenscoff:1980,Blaschke:2020acc}.
Hence, we presently subtract the static contribution and will add it at the end of our derivation, more precisely, we will add the well-known solution to the static problem at the very end so as not to clutter our equations in intermediate steps.
Focusing only on the dynamic part of the accelerating dislocation field, we presently replace \eqref{eq:transformedbound1a} with
\begin{align}
U_0\equiv \lim_{y\to0^+}\widetilde U_1(\lambda,y,s) &= -\frac{b}{2s}\int_0^\infty e^{s(\lambda x'-\eta(x'))}dx'
\,,\label{eq:transformedbound1b}
\end{align}
and for notational simplicity we drop the tilde below ($\widetilde U_1\to U_1$).
We furthermore assume that $c_{12}+c_{66}\ne0$, i.e. we do not include the so-called irregular hyperbolic case \cite{Markenscoff:1985a} in our discussion, as we are unaware of any slip systems that in practice would exhibit this property \cite{Blaschke:2021vcrit}.
The differential equations \eqref{eq:diffeq_transformed} can be rewritten in 4x4 matrix form as
\begin{align}
\left(\begin{matrix}
0 & \delta_{kj} \\
-s_{ki}\left(C_{i11j}\lambda^2-\rho\delta_{ij}\right)s^2 & s_{ki}\left(C_{i12j}+C_{i21j}\right)s\lambda
\end{matrix}\right)
\left(\begin{matrix}
U_j \\ \partial_2 U_j
\end{matrix}
\right)
=\partial_2\left(\begin{matrix}
U_k \\ \partial_2 U_k
\end{matrix}
\right)
\,,
\end{align}
where we defined the compliances as $s_{ki}C_{i22j}\equiv \delta_{kj}$, i.e.
\begin{align}
s_{11} &= \frac{c_{22}}{c_{22} c_{66} - c_{26}^{2}}\,, &
s_{12} &= s_{21}= \frac{-c_{26}}{c_{22} c_{66} - c_{26}^{2}}\,, &
s_{22} &= \frac{c_{66}}{c_{22} c_{66} - c_{26}^{2}}
\,. \label{eq:compliances}
\end{align}
Since we focus here on the regular hyperbolic case, we may assume that the eigenvalues of the so-defined 4x4 matrix ($\mu_m$ with $m=\pm1,\pm2$) are distinct \cite{Markenscoff:1987}.
Given these eigenvalues, we make the ansatz
\begin{align}
U_j(\lambda,y,s) &= \sum_{m} A_{jm}(\lambda,s) e^{-\mu_m sy}
\,. \label{eq:ansatzUj}
\end{align}
Plugging this ansatz into the differential equations \eqref{eq:diffeq_transformed} yields the determinantal equation
\begin{align}
\det\left(C_{i11j}\lambda^2 - \left(C_{i12j}+C_{i21j}\right)\lambda\mu_m + C_{i22j}\mu_m^2 - \rho\delta_{ij}\right)=0
\,, \label{eq:determinantal}
\end{align}
which may be used to calculate the $\mu_m(\lambda)$ by solving the following fourth order polynomial:
\begin{align}
0 &= 
\mu^4\left(c_{22}c_{66} - c_{26}^2\right) \nn\\
&\quad - 2\mu^3 \lambda \left(c_{22}c_{16} - c_{26}c_{12}\right) 
\nn\\
&\quad - \mu^2\left[c_{22}\left(\rho - \lambda^2 c_{11} \right) + c_{66}\left(\rho - \lambda^2c_{66} \right) + \lambda^2\left(c_{12}+c_{66}\right)^2 - 2\lambda^2c_{16}c_{26} \right] 
\nn\\
&\quad + 2 \mu\, \lambda \left[  c_{26}\left(\rho - \lambda^2 c_{11} \right)  +  c_{16}\left(\rho - \lambda^2c_{66}\right) + \lambda^2 c_{16}\left(c_{12}+c_{66}\right)\right]
\nn\\
&\quad + \left(\rho - \lambda^2 c_{11}\right)\left(\rho - \lambda^2c_{66} \right) - \lambda^4 c_{16}^2
\,. \label{eq:thequarticeqformu}
\end{align}
Note that $s$ factored out in this equation so that $\mu_m$ depends on $\lambda$ but not on $s$.
Finally, the asymptotic condition $\lim\limits_{y\to\infty}\pa_2U_i=0$ tells us that the sum over $m$ in the ansatz \eqref{eq:ansatzUj} above must only include the positive eigenvalues and Markenscoff argued in \cite{Markenscoff:1987} that because the slowness surface (whose equation coincides with the determinantal equation \eqref{eq:determinantal} above) is symmetric about the origin, there are presently two positive eigenvalues, $m=1,2$.
The corresponding eigenvectors are $\left(A_{1m},A_{2m},-\mu_msA_{1m},-\mu_msA_{2m}\right)$ where the $A_{im}$ is determined from
\begin{align}
\left(C_{i11j}\lambda^2 - \left(C_{i12j}+C_{i21j}\right)\lambda\mu_m + C_{i22j}\mu_m^2 - \rho\delta_{ij}\right)A_{jm}=0
\label{eq:eigveceq}
\end{align}
together with the boundary conditions which presently read
\begin{subequations}\label{eq:boundAjm}
\begin{align}
&A_{11}+A_{12} = U_1(\lambda,0,s)\equiv U_0\,,
\\
& c_{22}\left(\mu_1A_{21}+\mu_2A_{22}\right) - \lambda c_{12}(A_{11}+A_{12}) + c_{26}\left(\mu_1A_{11}+\mu_2A_{12} - \lambda (A_{21}+A_{22})\right)
=0\,.
\end{align}
\end{subequations}
Plugging the ansatz $A_{2m}=a_m A_{1m}$ into \eqref{eq:eigveceq}, we find for $a_m$:
\begin{align}
a_{m} = -\frac{\left(c_{11}\lambda^2 - 2c_{16}\lambda\mu_m + c_{66}\mu_m^2 - \rho\right)}{\left(c_{16}\lambda^2 - \left(c_{12}+c_{66}\right)\lambda\mu_m + c_{26}\mu_m^2\right)}
= -\frac{\left(c_{16}\lambda^2 - \left(c_{12}+c_{66}\right)\lambda\mu_m + c_{26}\mu_m^2\right)}{\left(c_{66}\lambda^2 - 2c_{26}\lambda\mu_m + c_{22}\mu_m^2 - \rho\right)}
\label{eq:amsol}
\end{align}
where the last equality follows from the fact that $\mu_m$ solves \eqref{eq:determinantal}.
The boundary conditions \eqref{eq:boundAjm} finally determine $A_{1m}$, and written in matrix form we presently have
\begin{align}
\left(\begin{matrix}
 1 & 1 \\
 c_{22}\mu_1 a_1 - \lambda c_{12} + c_{26}\left(\mu_1 - \lambda a_1\right)
 & c_{22}\mu_2a_2 - \lambda c_{12} + c_{26}\left(\mu_2 - \lambda a_2\right)
\end{matrix}\right)
\left(\begin{matrix}
A_{11} \\ A_{12}
\end{matrix}\right)
 = \left(\begin{matrix}
 	U_{0} \\ 0
 \end{matrix}\right)
\,.
\end{align}
Thus,
\begin{align}
A_{12} &= U_{0} - A_{11} \,, \nonumber\\
A_{11} &= \frac{-\left[c_{22}\mu_2a_2 - \lambda c_{12} + c_{26}\left(\mu_2 - \lambda a_2\right)\right]}{c_{22}(\mu_1 a_1-\mu_2a_2) + c_{26}\left(\mu_1 - \lambda a_1\right) - c_{26}\left(\mu_2 - \lambda a_2\right)}U_{0}
\label{eq:A1msol}
\end{align}
with $a_m$ given in \eqref{eq:amsol}.
Note that the coefficients $A_{im}(\lambda)$ do not depend on $s$; this will be important later when we derive the inverse Laplace transform.

\section{Cagniard-de Hoop method}
\ \\
In order to determine the displacement gradient field in real space and time, we need to apply the inverse Laplace transform $\cT^{-1}\{f\}(x) = \frac{1}{2\pi i}\int_{\epsilon-i\infty}^{\epsilon+i\infty}f(\lambda)e^{-s\lambda x}s d\lambda$ and integrate $\lambda$ along a line parallel to the imaginary axis.
This latter integral will not be carried out explicitly, but rather we want to rewrite it in a way that allows us to interpret this integral as a Laplace transform in time so that a subsequent inversion of the one sided Laplace transform $\cL\{u_i\}$ need not be carried out explicitly.

Thus, for each term in $U_i$ we interpret the following combination as a strictly positive time variable $\tau$ in order to apply the Cagniard-de Hoop method \cite{Cagniard:1939,DeHoop:1960,Freund:1973}:
\begin{align}
	\tau_m \equiv y \mu_m (\lambda) + (x - x') \lambda \ge 0
	\,. \label{eq:tauoflambda}
\end{align}
The reader is reminded that we presently restrict our calculation to the half plane $y\ge0$.
In order to be able to integrate $\tau$ over the positive real axis instead of over the imaginary $\lambda$ axis, one needs to study an integral over $\lambda$ over a closed path in complex space and to account for the residua of all enclosed poles.
This step requires knowledge of the locations of all poles in the expressions above, and hence knowledge of the roots $\mu_m(\lambda)$.
Note, that such poles occur only for transonic and supersonic dislocations, but not in the subsonic regime \cite{Markenscoff:1987}.
Furthermore, in passing from integration variable $\lambda$ to integration variable $\tau_m$, we need  the inverse of function \eqref{eq:tauoflambda}, i.e. $\lambda_m(\tau_m)$, as well as the Jacobian $\frac{d\lambda_m}{d\tau_m}$.
The inverted functions $\lambda_m$ appear in complex conjugate pairs which both need to be taken into account in order to integrate over a closed path \cite{Markenscoff:1987,Blaschke:2020acc}.
Using Cauchy's theorem we presently have in the subsonic regime:
\begin{align}
\cL\{u_j\} &= \cL\{u_j^\text{static}\} - \frac{b}{4\pi i}\int\limits_{\epsilon-i\infty}^{\epsilon+i\infty}d\lambda \, e^{-s\lambda x}\sum_{m=1}^2 \tilde{A}_{jm}(\lambda,s) e^{-\mu_m sy}\int\limits_0^\infty dx' e^{s(\lambda x'-\eta(x'))}
\nonumber\\
&= \cL\{u_j^\text{static}\} - \frac{b}{2\pi}\sum_{m=1}^2 \int_0^\infty\!\! dx'\, \Im\left[ \int_{\tau_m^\text{min}}^\infty d\tau_m \frac{d\lambda_m}{d\tau_m} \tilde{A}_{jm}(\lambda_m)e^{-s\tau_m} e^{-s\eta(x')}\right]
\end{align}
where $\tau_m^\text{min}=\lim\limits_{\lambda\to0}\tau_m(\lambda)$ and ${A}_{jm}=\tilde{A}_{jm}U_0$ is given in \eqref{eq:A1msol} with \eqref{eq:amsol}.
In the transonic and supersonic regimes, the expression above needs to be supplemented by appropriate residua for all enclosed poles in the integration path.
As discussed in earlier papers \cite{Markenscoff:1980,Blaschke:2020acc}, calculating $u_j$ directly is troublesome due to subtleties with respect to poles, and it is generally better to solve for its gradient.
Thus, taking derivatives with respect to $x$ and $y$ prior to passing from $\lambda$ to $\tau$, we find
\begin{subequations}
\begin{align}
\cL\{\partial_x u_j\}
&= \cL\{\partial_x u_j^\text{static}\} 
+ \frac{b}{2\pi}\sum_{m=1}^2 \int_0^\infty\!\! dx'\, \Im\left[ s \int_{\tau_m^\text{min}}^\infty d\tau_m \lambda_m \frac{d\lambda_m}{d\tau_m} \tilde{A}_{jm} e^{-s\tau_m} e^{-s\eta(x')}\right]
\\
\cL\{\partial_y u_j\}
&= \cL\{\partial_y u_j^\text{static}\} 
+ \frac{b}{2\pi}\sum_{m=1}^2 \int_0^\infty\!\! dx'\, \Im\left[ s \int_{\tau_m^\text{min}}^\infty d\tau_m \mu_m\frac{d\lambda_m}{d\tau_m} \tilde{A}_{jm} e^{-s\tau_m} e^{-s\eta(x')}\right]
\,.
\end{align}
\end{subequations}
Another important subtlety concerns the exchange of integrals over $\lambda$ and $x'$ prior to the change of variables, which is only permissible if both integrations converge absolutely; this is not the case in general and a remedy was put forward in the context of pure screw disloctions in Refs. \cite{Markenscoff:1980,Blaschke:2020acc}.
In particular, the exchange of integrals leads to poles on the slip plane at $y\to0$ which stem from the first two terms of a Taylor expansion of $\eta(x')$ around $x'=x$.
On the other hand, if one were to replace $\eta$ with its linear order Taylor expansion terms, the integral over $x'$ can be carried out analytically before changing integration variables:
\begin{align}
\int_0^\infty\!\! dx'\, e^{s\left[\lambda x' - \tilde\eta(x,x')\right]} =  \frac{e^{-s\left[\eta(x)-x\eta'(x)\right]}}{s\left(\eta'(x) - \lambda\right)}
\,.
\end{align}
In that case, $\tau$ will not depend on $x'$ (i.e. one defines \eqref{eq:tauoflambda} with $x'=0$) and only one integral over $\lambda$ (resp. $\tau_m$) is left.

To sum up:
In order to eliminate divergences on the slip plane in the $x'$ integration, we must add and subtract the dynamic term with $\eta(x')$ replaced by its linear order Taylor expansion $\tilde\eta\equiv \eta(x) + (x'-x)\eta'(x)$ with $\eta'(x)\equiv\sgn{x}\partial_x\eta(\abs{x})$ and $\eta(x)\equiv \sgn{x}\eta(\abs{x})$, see Ref. \cite{Blaschke:2020acc}.
Hence,
\begin{subequations}
\begin{align}
\cL\{\partial_x u_j\}
&= \cL\{\partial_x u_j^\text{static}\} 
+ \frac{b}{2\pi}\sum_{m=1}^2 \, \Im\left[ \int_{\tau_m^\text{min}}^\infty d\tau_m \lambda_m \frac{d\lambda_m}{d\tau_m} \tilde{A}_{jm} e^{-s\tau_m} \frac{e^{-s\left[\eta(x)-x\eta'(x)\right]}}{\left(\eta'(x) - \lambda_m\right)}\right]
\nonumber\\
&\quad\qquad + \frac{b}{2\pi}\sum_{m=1}^2 \int_0^\infty\!\! dx'\, \Im\left[ s \int_{\tau_m^\text{min}}^\infty d\tau_m \lambda_m \frac{d\lambda_m}{d\tau_m} \tilde{A}_{jm} e^{-s\tau_m} \left(e^{-s\eta(x')} - e^{-s\tilde\eta(x,x')}\right)\right]
\\
\cL\{\partial_y u_j\}
&= \cL\{\partial_y u_j^\text{static}\} 
+ \frac{b}{2\pi}\sum_{m=1}^2 \, \Im\left[ \int_{\tau_m^\text{min}}^\infty d\tau_m \mu_m\frac{d\lambda_m}{d\tau_m} \tilde{A}_{jm} e^{-s\tau_m} \frac{e^{-s\left[\eta(x)-x\eta'(x)\right]}}{\left(\eta'(x) - \lambda_m\right)}\right]
\nonumber\\
&\quad\qquad + \frac{b}{2\pi}\sum_{m=1}^2 \int_0^\infty\!\! dx'\, \Im\left[ s \int_{\tau_m^\text{min}}^\infty d\tau_m \mu_m\frac{d\lambda_m}{d\tau_m} \tilde{A}_{jm} e^{-s\tau_m}  \left(e^{-s\eta(x')} - e^{-s\tilde\eta(x,x')}\right)\right]
\,.
\end{align}
\end{subequations}
Considering the properties of the Laplace transform, where multiplication by $e^{-sT}$ corresponds to a translation in time $t\to t-T$ and multiplication by $s$ corresponds to a time derivative (modulo boundary terms which are zero here),
we can read off the solution:
\begin{subequations}\label{eq:generalsolution}
\begin{align}
\partial_x u_j
&= \partial_x u_j^\text{static}
+ \frac{b}{2\pi}\sum_{m=1}^2 \, \Im\left[ \Theta\left(t - \left[\eta(x)-x\eta'(x)\right] - \tau_m^\text{min}\right) \lambda_m \frac{d\lambda_m}{dt}  \frac{\tilde{A}_{jm}}{\left(\eta'(x) - \lambda_m\right)}\right]
\nonumber\\
&\quad + \frac{b}{2\pi}\partial_t \int_0^\infty\!\! dx' \sum_{m=1}^2 \Im\left[ \lambda_m \frac{d\lambda_m}{dt} \tilde{A}_{jm} \left(\Theta\left(t - \eta(x') - t^\text{min}\right) - \Theta\left(t - \tilde\eta(x,x') - t^\text{min}\right)\right)\right]
\\
\partial_y u_j
&= \partial_y u_j^\text{static}
+ \frac{b}{2\pi}\sum_{m=1}^2 \, \Im\left[ \Theta\left(t - \left[\eta(x)-x\eta'(x)\right] - \tau_m^\text{min}\right) \mu_m\frac{d\lambda_m}{dt} \frac{ \tilde{A}_{jm} }{\left(\eta'(x) - \lambda_m\right)}\right]
\nonumber\\
&\quad + \frac{b}{2\pi}\partial_t\int_0^\infty\!\! dx'\sum_{m=1}^2 \Im\left[ \mu_m\frac{d\lambda_m}{d\tau_m} \tilde{A}_{jm} \left(\Theta\left(t - \eta(x') - t^\text{min}\right) - \Theta\left(t - \tilde\eta(x,x') - t^\text{min}\right)\right)\right]
\,,
\end{align}
\end{subequations}
where $\lambda_m$ depends on the appropriately shifted time $\tau=t - \left[\eta(x)-x\eta'(x)\right]$, $\tau = t - \eta(x')$, or $\tau = t - \tilde\eta(x,x')$, i.e. matching in each term the according part of the argument of the step function.

\section{Special cases: constant velocity and constant acceleration rate}

\begin{figure}[ht]
\centering
\includegraphics[trim=0 0.cm 1.8cm 0,clip,width=0.25\textwidth]{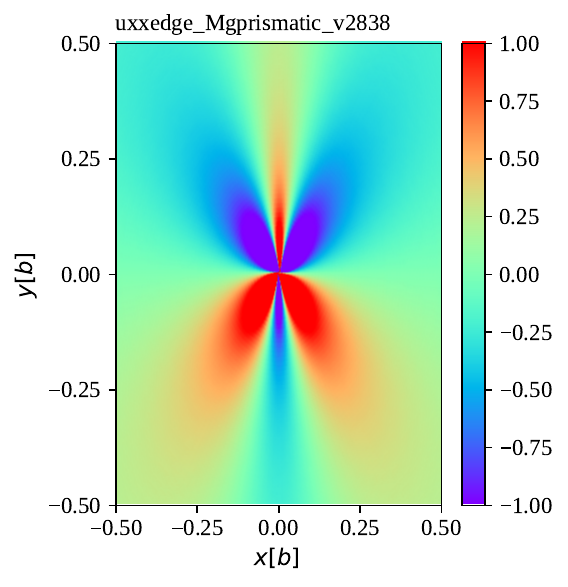}%
\includegraphics[trim=1.75cm 0 0.05cm 0,clip,width=0.25\textwidth]{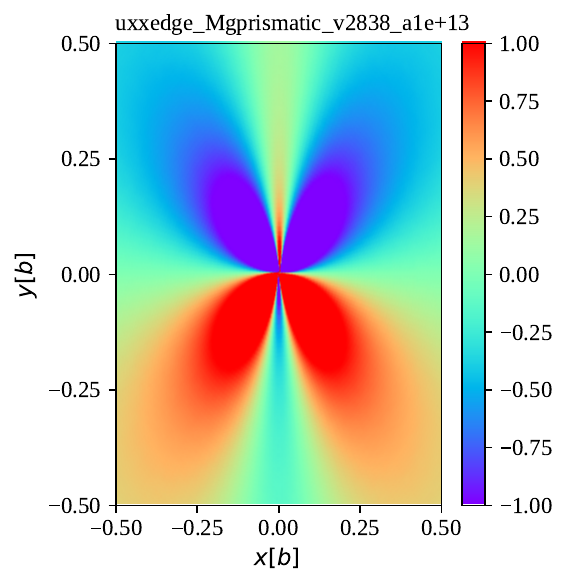}%
\includegraphics[trim=0 0.cm 1.8cm 0,clip,width=0.25\textwidth]{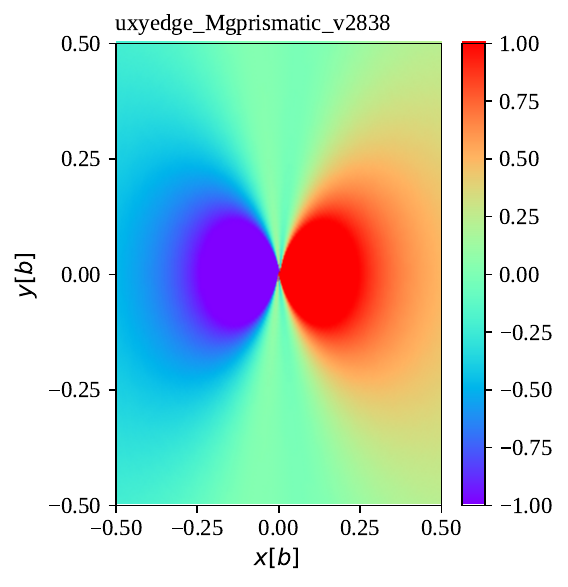}%
\includegraphics[trim=1.75cm 0 0.05cm 0,clip,width=0.25\textwidth]{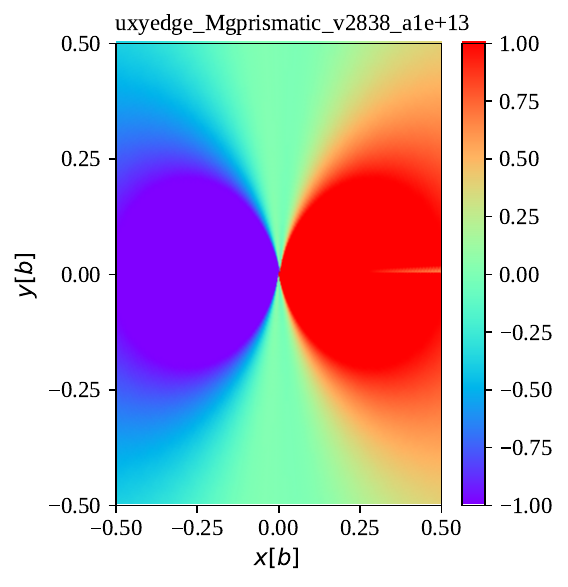}
\includegraphics[trim=0 0.cm 1.8cm 0,clip,width=0.25\textwidth]{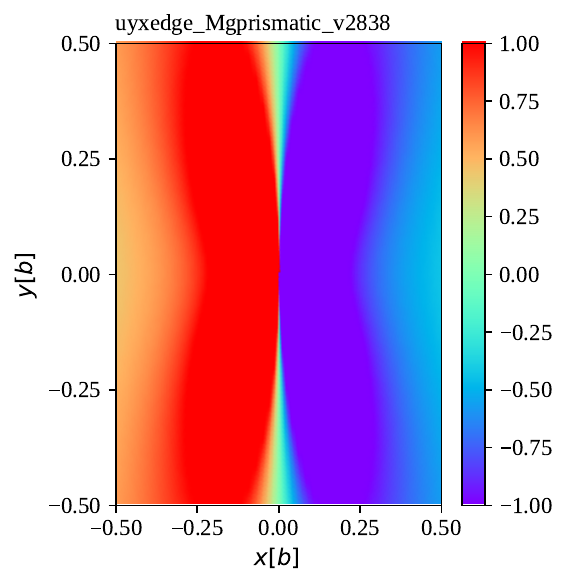}%
\includegraphics[trim=1.75cm 0 0.05cm 0,clip,width=0.25\textwidth]{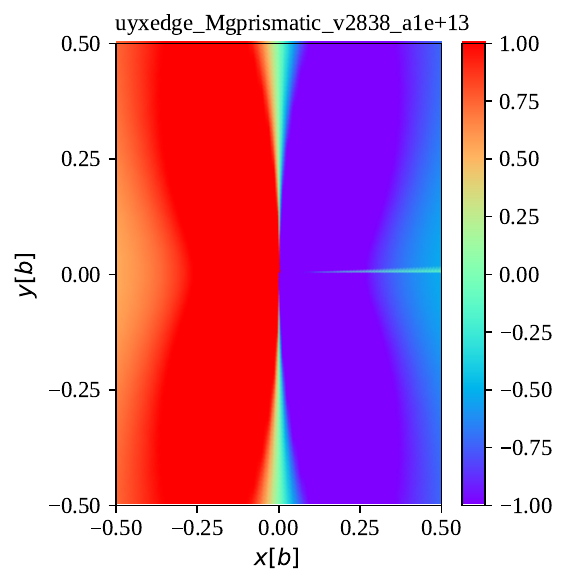}%
\includegraphics[trim=0 0.cm 1.8cm 0,clip,width=0.25\textwidth]{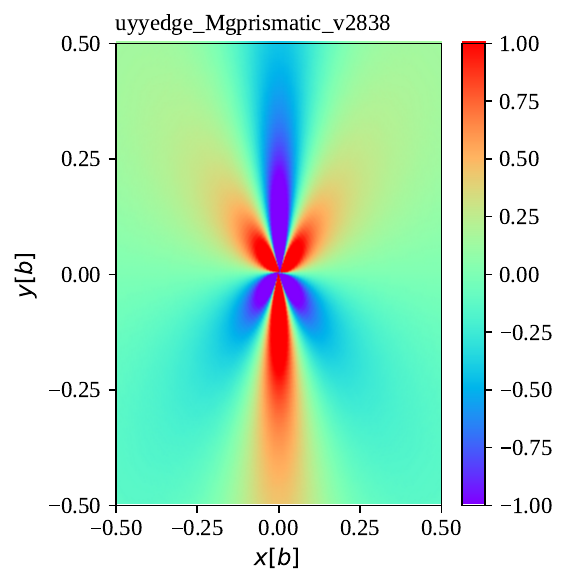}%
\includegraphics[trim=1.75cm 0 0.05cm 0,clip,width=0.25\textwidth]{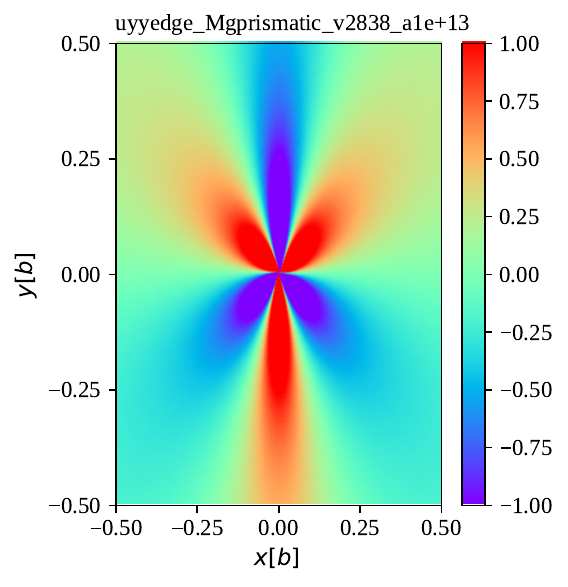}
\caption{We show $\partial_i u_{j}$ at dislocation velocity
$v=2.838\txt{km/s}$
for hcp Mg and prismatic slip ($\rho=1.74$g/ccm, $b=3.21$\r{A}, $c_{11}=59.5$GPa, $c_{12}=26.12$GPa, $c_{13}=21.805$GPa, $c_{33}=61.55$GPa, and $c_{44}=16.35$GPa, see \cite{CRCHandbook}).
This velocity corresponds to roughly 92\% of the critical velocity.
All plots are centered at the dislocation core, showing the plane perpendicular to the dislocation line in units of a Burgers vector.
On the left of each pair of plots, we show the steady state-solution \cite{Bacon:1980} and on the right we show the full solution for constant acceleration \eqref{eq:generalsolution} with \eqref{eq:constacc} and $a=1\times10^{13}$m/s$^2$ at time $t_v= v/a = 2.838\times10^{-10}$s needed to reach velocity $v$.
At this point, the dislocation has traveled a distance of $0.4$ microns.
We see that the changes in the dislocation displacement gradient due to the inclusion of acceleration lead to a slight enhancement.}
\label{fig:comparesteadystate}
\end{figure}

\begin{figure}[ht]
\centering
\includegraphics[trim=0 0.cm 1.8cm 0,clip,width=0.25\textwidth]{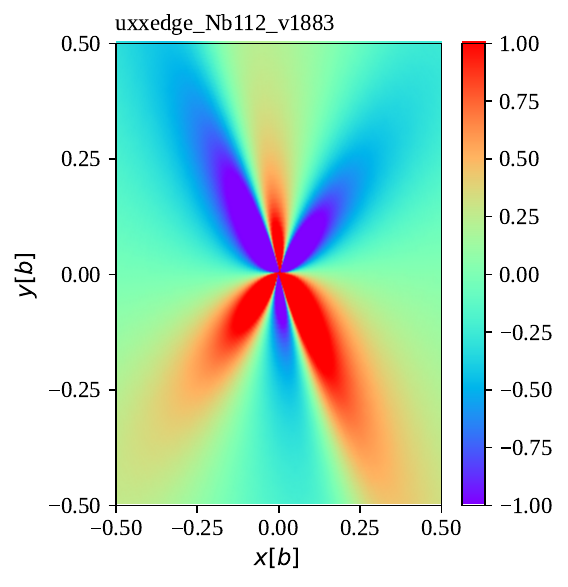}%
\includegraphics[trim=1.75cm 0 0.05cm 0,clip,width=0.25\textwidth]{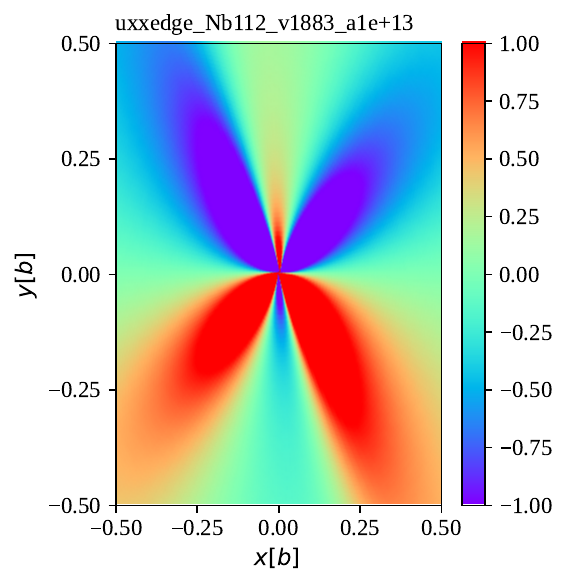}%
\includegraphics[trim=0 0.cm 1.8cm 0,clip,width=0.25\textwidth]{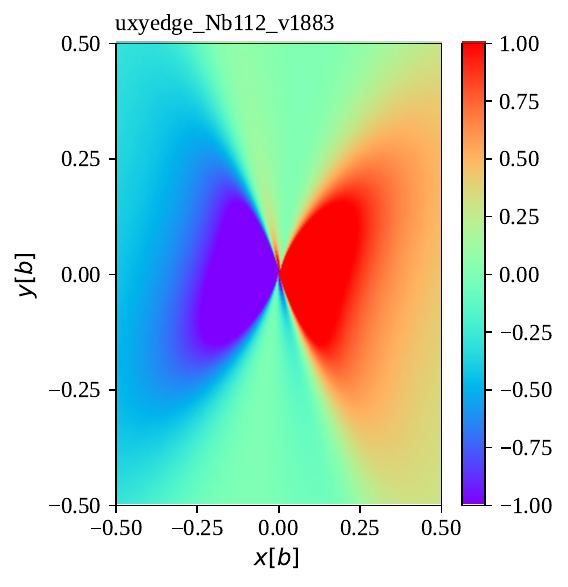}%
\includegraphics[trim=1.75cm 0 0.05cm 0,clip,width=0.25\textwidth]{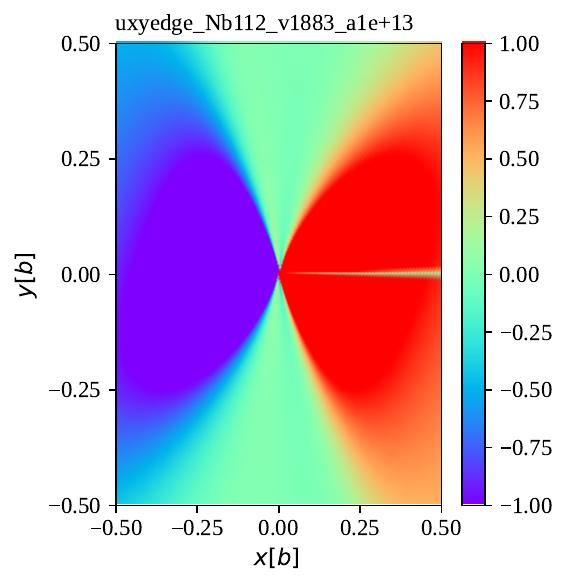}
\includegraphics[trim=0 0.cm 1.8cm 0,clip,width=0.25\textwidth]{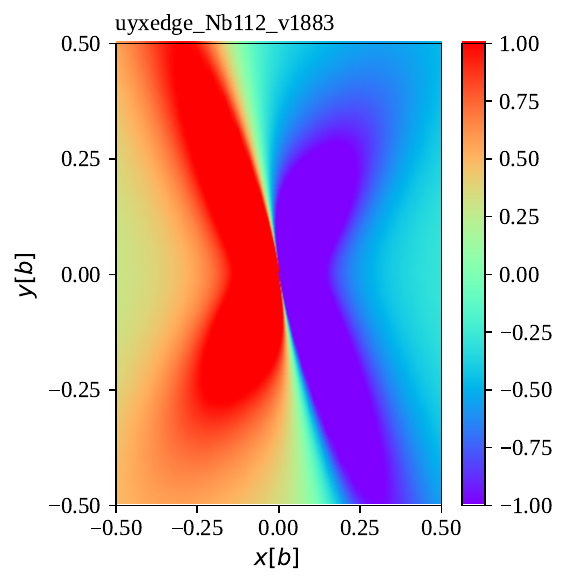}%
\includegraphics[trim=1.75cm 0 0.05cm 0,clip,width=0.25\textwidth]{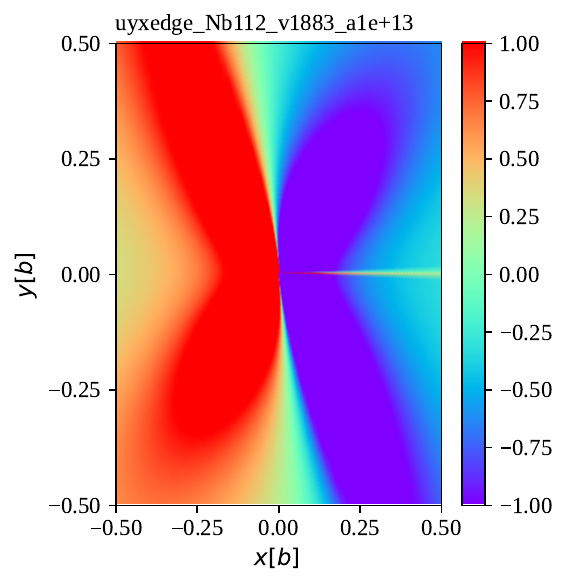}%
\includegraphics[trim=0 0.cm 1.8cm 0,clip,width=0.25\textwidth]{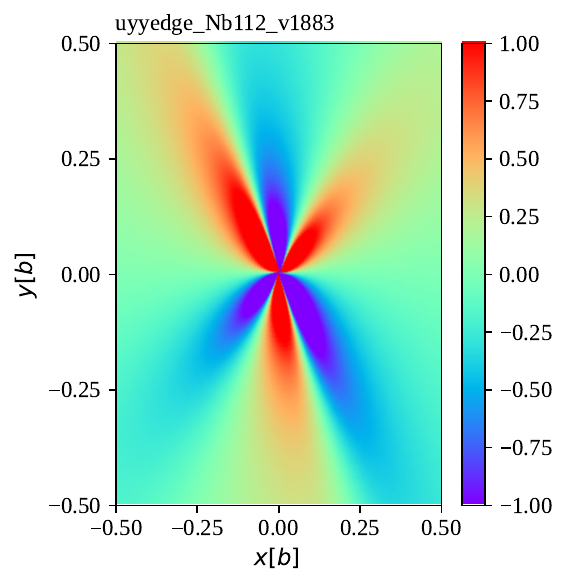}%
\includegraphics[trim=1.75cm 0 0.05cm 0,clip,width=0.25\textwidth]{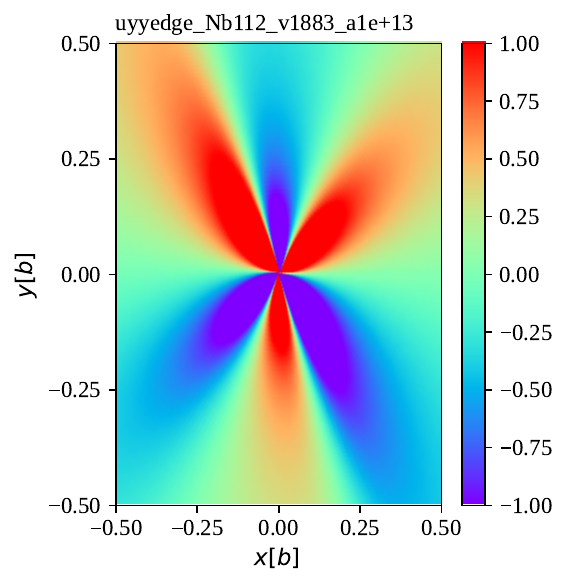}
\caption{We show $\partial_i u_{j}$ at dislocation velocity
$v=1.883\txt{km/s}$
for an edge dislocation in bcc Nb gliding on a 112 slip plane ($\rho=8.57$g/ccm, $b=2.86$\r{A}, $c_{11}=246.5$GPa, $c_{12}=134.5$GPa, and $c_{44}=28.73$GPa, see \cite{CRCHandbook}).
This velocity corresponds to roughly 90\% of the critical velocity.
All plots are centered at the dislocation core, showing the plane perpendicular to the dislocation line in units of a Burgers vector.
We compare the steady state-solution \cite{Bacon:1980} with the full solution for constant acceleration \eqref{eq:generalsolution} with \eqref{eq:constacc} and $a=1\times10^{13}$m/s$^2$ at time $t_v= v/a = 1.883\times10^{-10}$s needed to reach velocity $v$.
At this point, the dislocation has traveled a distance of $\sim0.18$ microns.
We see that the changes in the dislocation displacement gradient due to the inclusion of acceleration lead to a slight enhancement.}
\label{fig:comparesteadystate_Nb}
\end{figure}

\begin{figure}[ht]
\centering
\includegraphics[width=0.33\textwidth]{uyxedge_Mgprismatic_v2838_a1e+13.pdf}%
\includegraphics[width=0.33\textwidth]{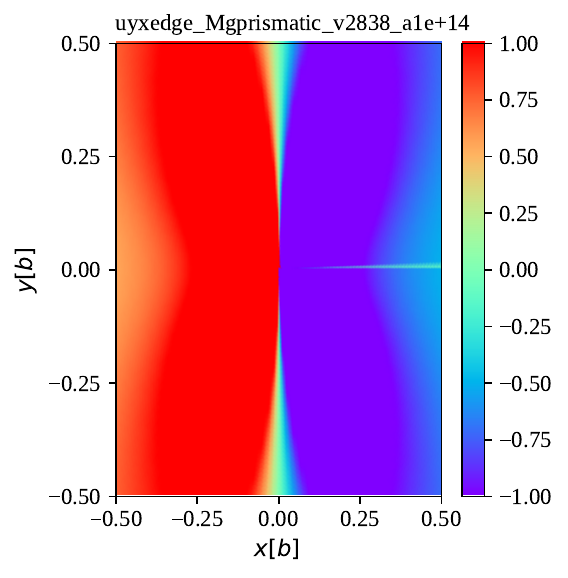}%
\includegraphics[width=0.33\textwidth]{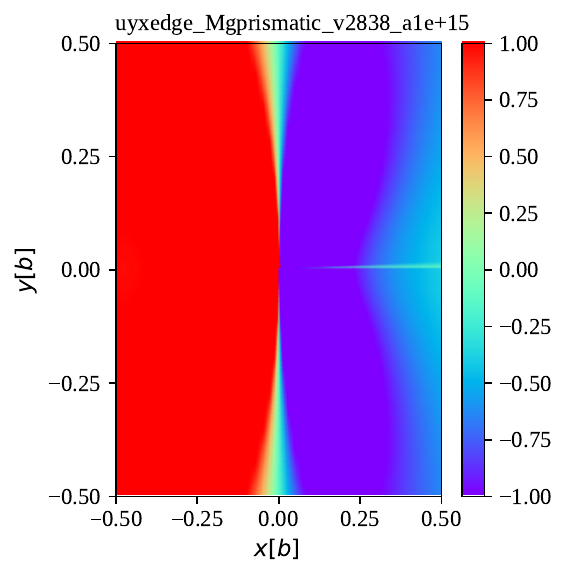}%
\caption{We compare $\partial_yu_{x}$ in Mg (prismatic slip) at dislocation velocity $v=2.838\txt{km/s}$ for different acceleration rates.}
\label{fig:compareaccelerationrate}
\end{figure}

The simplest case one can study within the present solution is a dislocation initially at rest which suddenly starts moving at constant velocity $v$ at time $t\ge0$.
As discussed previously in the context of pure screw dislocations in \cite{Blaschke:2020acc}, this ``jump'' in velocity is unphysical, but in the large time limit the solution must tend to the well-known steady state solution, thus providing us with a consistency check.
The assumption of constant dislocation velocity at $t\ge0$ leads to the following simplifications:
\begin{align}
\eta(x) &= \frac{x}{v}
\,, & \eta'(x)&=\frac{1}{v}
\,,& t - \left(\eta(x)-x\eta'(x)\right) &= t
\,, &
\tilde{\eta} &= \frac{x'}{v}=\eta(x')
\,. \label{eq:constvel}
\end{align}
Due to the last equality, the second and fourth lines within \eqnref{eq:generalsolution} (i.e. the terms containing the time derivative and the integral over $x'$) vanish identically for a dislocation moving at constant velocity.

The simplest physical case within the present dynamic solution, follows from the assumption that the dislocation is at rest at time $t<0$ and starts to accelerate at a constant rate $a$ from time $t\ge0$.
Then $l(t)=\frac{a}{2}t^2>0$ and hence \cite{Blaschke:2020acc}
\begin{align}
\eta(x) &= \sgn{x}\sqrt{\frac{2\abs{x}}{a}}
\,, & \eta'(x)&
=\frac{\eta(x)}{2x}
\,,& t - \left(\eta(x)-x\eta'(x)\right)
&= t - \frac{1}{2}\eta(x)
\,, &
\tilde{\eta} &= 
\frac{1}{2}\left(1+\frac{x'}{x}\right)\eta(x)
\,. \label{eq:constacc}
\end{align}
The velocity at time $t$ is given by $v(t)=at$ and the transition from subsonic to transonic happens when $t=v_\txt{lim}/a$, where $v_\txt{lim}$ is the lowest limiting velocity whose value can easily be computed using the review article \cite{Blaschke:2021vcrit} and/or the open source code \cite{pydislocdyn}.

We have implemented this constant acceleration rate case in Python, using a combination of symbolic (sympy) calculations and numerical methods, and have integrated it into the code PyDislocDyn \cite{pydislocdyn}.
The general strategy is as follows:
The material's tensor of elastic constants is rotated into coordinates where the dislocation line is parallel to the $z$ direction, the slip plane normal points in the $y$ direction and the edge dislocation accelerates from rest in the $x$ direction at rate $a$.
We then calculate the time $t_1$ at which the accelerating dislocation reaches a user-specified target velocity, as well as the position of the dislocation core at that time in order to shift the $x$ coordinate such that the dislocation core resides at the origin at time $t_1$.
We use sympy to calculate the four solutions $\mu(\rho/\lambda^2)$ from Eq. \eqref{eq:thequarticeqformu} after plugging in numerical values for all (rotated) elastic constants and the material density, i.e. $\lambda$ is the only unknown.
For each of these 4 solutions, we determine $\tau(\lambda)$ and its derivative, and the resulting sympy expressions are subsequently `lambdified', i.e. converted into functions of $\lambda$, $x$, and $y$.
We then loop over all points $x,y$ we wish to determine the displacement gradient for.
At a given point $x,y$, function $\tau$ depends only on $\lambda$, and since we are interested in one snapshot in time (meaning we know $\tau$), 
we can numerically determine $\lambda(\tau)$; note that $\lambda$ is a complex number and we use mpmath's recommended root finding method (the Muller method).
This step constitutes the bottleneck of our implementation, i.e. calculating the dislocation field for accelerating edge dislocations is orders of magnitude slower than for screw dislocations which were discussed in \cite{Blaschke:2020acc}.
Once we have $\lambda$, we determine $\mu(\lambda)$ and the Jacobian $1/\left(\frac{d\tau}{d\lambda}\right)$.
At this point we have 4 sets of $\lambda$, $\mu(\lambda)$, but only 2 satisfy the asymptotic condition $\lim\limits_{y\to\infty}\pa_2U_i=0$.
Markenscoff \cite{Markenscoff:1987} determined that the imaginary parts of $\lambda$ and $\mu/\lambda$ must have opposite signs for positive $y$, and we drop the other two solutions to $\lambda$.
The remaining two sets of $\lambda, \mu(\lambda)$ are plugged into \eqref{eq:amsol} and \eqref{eq:A1msol}, and subsequently into the first (i.e. leading) dynamic terms of \eqref{eq:generalsolution}.
The static part is computed with the well-known Stroh / integral method \cite{Bacon:1980}.
The time-derivative term in \eqref{eq:generalsolution} can be neglected for constant acceleration rates.

Figure \ref{fig:comparesteadystate} shows the edge dislocation field at the example of hcp Mg for prismatic slip and compares the accelerating field to the steady state field.
Figure \ref{fig:comparesteadystate_Nb} shows the edge dislocation field at the example of bcc Nb for the 112 slip planes and compares the accelerating field to the steady state field.
In contrast to the previous example, edge dislocations on 112 slip planes of bcc metals have a non-vanishing (rotated) elastic constant $c'_{26}$, and thus represent a more general case than the former.
Both examples show some enhancement of the dislocation displacement gradient field for moderate acceleration rates of $a\sim10^{13}$ m/s$^2$ typical for flyer plate impact scenarios \cite{Blaschke:2021impact}, albeit maintaining the shape of the steady state solution for the most part.
Only for very extreme acceleration rates do we start to see more significant deviations as illustrated in Fig. \ref{fig:compareaccelerationrate} at the example of Mg.
Note that the numerical accuracy of the accelerating edge solution in its current implementation is limited by the accuracy of the (complex) root finding algorithm.

Furthermore, we confirm (numerically) that the divergence at a `critical' dislocation velocity (which separates the subsonic from the transonic regime), persists for general accelerating edge dislocations with vanishing core size, consistent with previous work on the isotropic limit \cite{Markenscoff:2008} as well as the accelerating screw dislocation in anisotropic crystals \cite{Blaschke:2020acc}.

\section{The isotropic limit}

The following simplifications apply in the isotropic limit:
$c_{22}=c_{11}=c_{12}+2c_{44}$, $c_{66}=c_{44}$, and $c_{16}=0=c_{26}$, as well as $s_{11}=1/c_{11}=s_{22}$ and $s_{12}=0=s_{21}$ within \eqref{eq:compliances}.
Hence, \eqnref{eq:thequarticeqformu} simplifies to
\begin{align}
0&=\mu^4c_{11}c_{44}
 - \mu^2\left[c_{11}\left(\rho - \lambda^2 c_{11} \right) + c_{44}\left(\rho - \lambda^2c_{44} \right) + \lambda^2\left(c_{12}+c_{44}\right)^2 \right]
 + \left(\rho - \lambda^2 c_{11}\right)\left(\rho - \lambda^2c_{44} \right)
\,,
\end{align}
where $c_{11}=c_{12}+2c_{44}$, and solutions $\mu_m$ are found to be
\begin{align}
\mu_1 &= \pm\sqrt{\frac{\rho}{c_{44}}-\lambda^2}
\,,&
\mu_2 &= \pm\sqrt{\frac{\rho}{c_{11}}-\lambda^2}
\,.
\end{align}
In both cases, only one of the two signs must be considered, namely convergence of \eqref{eq:ansatzUj} requires that 
the real part of $\mu_m$ has the same sign as $y$.
For positive $y$ this mean that $\Im(\lambda)>0$ implies $\Im(\mu_m/\lambda)<0$ and vice versa \cite{Markenscoff:1987}.

Coefficients $A_{im}$ simplify to
\begin{align}
\tilde{A}_{11} &= \frac{\lambda c_{12} - c_{11}\mu_2a_2}{c_{11}(\mu_1 a_1-\mu_2a_2)}
\,, &
\tilde{A}_{12} &= 1 - \tilde{A}_{11} \,, &
\tilde{A}_{2m} &= a_m \tilde{A}_{1m}\,,\nn\\
a_m &= \frac{\left(c_{11}\lambda^2 + c_{44}\mu_m^2 - \rho\right)}{\left(c_{12}+c_{44}\right)\lambda\mu_m }
= \frac{\left(c_{12}+c_{44}\right)\lambda\mu_m}{\left(c_{44}\lambda^2 + c_{11}\mu_m^2 - \rho\right)}
\,,\label{eq:AimIso}
\end{align}
with $c_{11}=c_{12}+2c_{44}$.

The definition of $\tau_m$ (with $x'=0$) then yields
\begin{align}
\lambda^\pm_m(\tau) &= \frac{\tau}{r^2}\left(x\pm i y \sqrt{1-\frac{r^2}{c_m^2\tau^2}}\right)
\,,\nn\\
\mu_m^\pm &= \frac1y\left(\tau - x\lambda^\pm_m\right)
= \frac{\tau}{r^2}\left(y\mp ix\sqrt{1-\frac{r^2}{c_m^2\tau^2}}\right)
\,,\nn\\
\frac{d\lambda^\pm}{d\tau} &= \frac{1}{r^2}\left(x \pm iy \frac{1}{\sqrt{1-\frac{r^2}{c_m^2\tau^2}}}\right)
=\frac{\pm i \mu^\pm_m}{\tau\sqrt{1-\frac{r^2}{c_m^2\tau^2}}}
\,,
\end{align}
with $r^2\equiv x^2+y^2$ and the short-hand notation $c_1\equiv \ct=\sqrt{c_{44}/\rho}$ and $c_2\equiv  \cl=\sqrt{c_{11}/\rho}$ for the transverse (T) and longitudinal (L) sound speeds.
This special case was discussed in Ref. \cite{Markenscoff:1981}.

If we assume a constant dislocation velocity from time $t>0$, i.e. $\eta(x)=x/v$ and take the limit of $t\to \infty$ after translating our coordinates to move with the dislocation (i.e. replacing $x=x'+vt$, $r^2=(x'+vt)^2+y^2$ everywhere prior to taking the limit, see \cite{Blaschke:2020acc}), we recover the well-known steady-state solution for an edge dislocation in an isotropic medium \cite{Eshelby:1949,Blaschke:2019Bpap}:
\begin{subequations}\label{eq:isosteadysolution}
\begin{align}
\partial_x u_x^\text{iso,steady}&=\frac{-b y}{\pi  \bt ^2 }\left(\frac{1/\gl }{\left((x-t v)^2 +y^2/\gl ^2\right)}-\frac{\left(1-\frac{\bt ^2}{2}\right) /\gt }{\left((x-t v)^2 +y^2/\gt ^2\right)}\right)
\,,\\
\partial_y u_x^\text{iso,steady}&=\frac{b (x-t v)}{\pi  \bt ^2 }\left(\frac{1/ \gl }{ \left((x-t v)^2 +y^2/\gl ^2\right)}-\frac{\left(1-\frac{\bt ^2}{2}\right)/ \gt }{\left((x-t v)^2 +y^2/\gt ^2\right)}\right)
\,,\\
\partial_x u_y^\text{iso,steady}&=\frac{b (x-t v)}{\pi  \bt ^2 }\left( \frac{1/\gl }{\left((x-t v)^2 +y^2/\gl ^2\right)}-\frac{\gt \left(1-\frac{\bt ^2}{2}\right) }{ \left((x-t v)^2 +y^2/\gt ^2\right)}\right)
\,,\\
\partial_y u_y^\text{iso,steady}&=\frac{b y}{\pi  \bt ^2 }\left(\frac{1/\gl ^3}{ \left((x-t v)^2 +y^2/\gl ^2\right)}-\frac{\left(1-\frac{\bt ^2}{2}\right)/ \gt }{\left((x-t v)^2 +y^2/\gt ^2\right)}\right)
\,.
\end{align}
\end{subequations}

\section{Conclusion}

In this paper, we have presented and discussed the full solution to the differential equations for an accelerating edge dislocation in a general anisotropic crystal in the subsonic regime.
Taking the formal solution of Ref. \cite{Markenscoff:1987} one step further, we have derived the edge dislocation displacement gradient field using a combination of analytical and numerical methods.
Our python implementation is included in version 1.2.7 of the code PyDislocDyn \cite{pydislocdyn}.
Two examples were illustrated in Figs. \ref{fig:comparesteadystate} and \ref{fig:comparesteadystate_Nb} showing that the dislocation strain field is slightly enhanced in the accelerating case, at least for typical dislocation acceleration rates of $a\sim10^{13}$ m/s$^2$ \cite{Blaschke:2021impact}, though still similar enough to the steady-state solution (except for extreme conditions such as very high acceleration rates and velocities near the limiting velocity), so that in most larger simulations it makes more sense to use the (several orders of magnitude) faster-to-compute steady state solution.
The transonic regime of the accelerating edge dislocation as well as accelerating mixed dislocations are left for future work.

\subsection*{Acknowledgements}
\noindent
We thank the anonymous referees for their valuable comments.

Research presented in this article was supported by the Laboratory Directed Research and Development program of Los Alamos National Laboratory under project number 20210826ER.
Furthermore, the authors are grateful for the support of the Materials project within the Advanced Simulation and Computing, Physics and Engineering Models Program of the U.S. Department of Energy under contract 89233218CNA000001 in the final stages of this work.

\bibliographystyle{utphys-custom}
\bibliography{dislocations}

\end{document}